 \documentclass[preprint2]{aastex}

\usepackage{natbib}
\usepackage{textcomp}
\usepackage{graphicx}
\newcommand{\un}[1]{\mbox{ \rmfamily #1}}
\newcommand{\unp}[1]{\mbox{\rmfamily #1}}
\newcommand{\unrho}[0]{\mbox{ \rmfamily kg} / \mbox{\rmfamily m} ^3}

\newcommand{\fig}[1]{Fig.~\ref{#1}}
\newcommand{\tab}[1]{Table~\ref{#1}}
\newcommand{\undeg}{\mbox{\textdegree}}

\newcommand{\unsec}{''}

\newcommand{\fo}{Find\_Orb}

\slugcomment{Submitted to ApJL}

\shorttitle{Radiation pressure on 2011~MD}
\shortauthors{Micheli, Tholen}

\begin{document}


\title{Radiation pressure detection and density estimate for 2011~MD}


\author{Marco Micheli, David J. Tholen, Garrett T. Elliott}
\affil{Institute for Astronomy, University of Hawaii, 2680 Woodlawn Dr., Honolulu, HI 96822, USA}
\email{micheli@ifa.hawaii.edu, tholen@ifa.hawaii.edu, gte@ifa.hawaii.edu}

\begin{abstract}
We present our astrometric observations of the small near-Earth object 2011~MD ($H \sim 28.0$), obtained after its very close fly-by to Earth in June 2011. Our set of observations extends the observational arc to $73$ days, and together with the published astrometry obtained around the Earth fly-by allows a direct detection of the effect of radiation pressure on the object, with a confidence of $5\sigma$. The detection can be used to put constraints on the density of the object, pointing to either an unexpectedly low value of $\rho = (640\pm330) \unrho$ ($68\%$ confidence interval) if we assume a typical probability distribution for the unknown albedo, or to an unusually high reflectivity of its surface. This result may have important implications both in terms of impact hazard from small objects and in light of a possible retrieval of this target.
\end{abstract}





\section{Introduction}

The small near-Earth asteroid 2011~MD was discovered on 2011 June 22 by the Lincoln Near-Earth Asteroid Research (LINEAR) survey in New Mexico, USA \citep{2011MPEC....M...23B}. Within 24 hours of discovery it was obvious that the object was going to have an extremely close approach to Earth in a few days, at about $18\,700 \un{km}$ from the Earth's center ($12\,300 \un{km}$ from its surface, flying over the Southern hemisphere).

Around its closest approach the object's magnitude peaked at about $V=11$, and it remained brighter than $V=19$ for four days before and after the peak. As a result, more than $1500$ individual astrometric positions were obtained and reported to the Minor Planet Center (MPC) in a period of less than $8$ days. However, the object rapidly became faint while receding from Earth, and no further observations were reported after 2011 July 3, only $11$ days after discovery.\\

Around that time we realized that the object was still fading at a reasonably slow rate of less than $0.5$ magnitudes per week, and we would have the capability to observe it for at least two more months using the telescopes to which we have access on Mauna Kea. We were able to obtain astrometric positions of the object on $5$ nights in August and early September 2011, therefore extending the observed arc on the object from $11$ to $73$ days, or about a factor of $6.5$.

In this work we present these observations, together with an accurate analysis of the object's dynamics made possible by this extended observational arc. We also discuss the implications of this result on the object's physical properties.

\subsection{Previous work}

The case of 2011~MD shares some resemblance with other very small NEOs observed in the past. Of about $200$ known small objects in this size range (diameter around or below $10 \un{m}$), only a handful remained observable from the ground for more than a few days, because of their intrinsic faintness. Only some peculiar characteristic of the close approach can allow for an extended observability window, enough to characterize their dynamical behavior in good detail.

The first example of one such object was probably 2006~RH120, an even smaller NEO that was temporarily captured in Earth orbit in 2006-2007 \citep{2008MPEC....D...12B,2009A&A...495..967K}. In that case, the long orbital phase allowed for 9 months of almost continuous observations from the ground. A second case was 2009~BD, that happened to have two very close approaches with Earth (and a couple of more distant but observable ones) in less than 3 years \citep{2009MPEC....B...14B,2012NewA...17..446M}. We recently presented our observational data and analysis of a third such object, 2012~LA \citep{2012MPEC....L...06H,2013Icar..226..251M}
All these objects, together with 2011~MD, share the property of having very Earth-like orbits, with very modest eccentricities and inclinations. As a result, they usually have very low relative orbital velocity with respect to our planet ($\Delta v < 4 \un{km}/\unp{s}$), making their close encounters last unusually long. This same property also implies that these same objects are also among the easiest to fly-by or rendezvous with a spacecraft launched from Earth; together with the small size, this makes them plausible candidates for a Asteroid Robotic Retrieval Mission (ARRM) such as the one currently under study by NASA.

However, the 2011~MD case is peculiar because it had only a single and short close encounter with Earth, and it was therefore observable for a much shorter timespan. Furthermore, being discovered only around the time of close approach, only the second half of the observability window was available.

\section{Methods}

One of the goals of our observational campaign on 2011~MD was to obtain enough astrometric information to detect non-gravitational forces acting on the object. However, the short observational arc posed additional challenges compared with previous cases, such as that presented in \citet{2012NewA...17..446M}.

The first and most obvious requirement is to obtain the highest possible signal to noise ratio during the observations, down to a magnitude of approximately $V=24$ at the end of the observable arc. For all our observations we used the University of Hawaii $2.24 \un{m}$ telescope atop Mauna Kea, equipped with a Tektronix 2048 CCD camera. All our observations were unfiltered, to maximize the SNR and improve the quality of the astrometry.

The second and equally fundamental step is to ensure that the highest possible astrometric quality is obtained from each image. We used custom software tools that are capable of performing high-precision astrometry on fields with significant trailing of the reference stars, as is the case in all our non-sidereally tracked exposures.
An accurate description of the techniques used in this work can be found in \citet{2013AcAau..87..147T}. It is worth noting that our astrometry presented in this work is referenced to the PPMXL catalog \citep{2010AJ....139.2440R}, currently believed to be among the least biased astrometric catalogs available, at least until a catalog from the Gaia mission will become available in the near future.\\

The choice of an appropriate catalog may be sufficient to minimize the possible astrometric biases of our own measurements. However, this work is based on the complete observed arc for 2011~MD, including more than $1500$ positions from other observatories, retrieved through the Minor Planet Center archive. For these positions, we do not have control on the catalog used, and it is possible that catalog biases are reflected in the astrometry. To minimize this effect we used a zone-specific debiasing following \citet{2010Icar..210..158C}, and applied the appropriate corrections to each coordinate before using the astrometry in our dynamical analysis.

A further important detail of a high-precision astrometric analysis is the use of an appropriate weighting procedure, based on the knowledge of an error bar associated with each astrometric position.
For our observations, a formally computed error bar is available as an output of our astrometry software, computed under the assumptions explained in \citet{2013AcAau..87..147T}. The error components from the astrometric solution and centroiding accuracy are directly estimated by the software, while the contribution from an unmodeled catalog bias has been conservatively estimated at $0.05\unsec$ for this analysis.
Unfortunately, positions from other sources outside our control usually don't have this information. In our analysis these missing error bars were replaced with the station-specific error values used by the NEODyS website\footnote{\url{newton.dm.unipi.it/\~{}neodys2/mpcobs/2011MD.rwo}. All URL-based references are to be intended as ``last accessed'' on 2014-04-12.}, which are known to be conservative in most cases because of a safety factor introduced to take into account unmodeled correlations; for the sake of a conservative analysis, we decided to maintain this safety factor in our work.

It is important to point out here that the 2011~MD dataset has an additional complication: a few stations reported an extremely high number of astrometric observations in a single night. For example, the Barred Owl Observatory (IAU code I27) reported $968$ positions to the MPC on the single night of 2011 June 27. Many other stations reported more than $10$ positions on at least one night. These large sets of observations are extremely dangerous for a dynamical analysis, because any error source or bias specific to that station will dominate the global dataset, introducing correlations in the raw data that cannot easily be accounted for. This is especially true for a fast-moving object as 2011~MD, where even a small clock error can cause a systematic residual in all positions reported from a station.
To prevent this effect we decided to down-weight every observatory that reported $N > 4$ observation in a single night\footnote{The choice of $4$ observations is based in part on the MPC rules, that discourage collecting ``many more than three observations per objects [\textit{sic}] per night'' (quoted from \url{www.minorplanetcenter.net/iau/info/Astrometry.html}).}. The new weight is computed multiplying the station-specific error bar discussed above by a factor of $\sqrt{N/4}$. As a result, if a set of positions from a single station in a single night contains more than $4$ observations, its total weight in the final orbital solution is the same as if they reported only $4$ positions. Using this approach we avoid the arbitrary rejection of some datapoints, while at the same time keeping the dataset mostly free from station-specific biases.

One final very important step is necessary to ensure that our astrometric dataset is cleaned of any possible source of systematic errors. We need to reject possible outliers, with the use a deterministic and statistically solid algorithm. In this work we use again our implementation of the Peirce criterion \citep{1852AJ......2..161P}, as presented in \citet{2012NewA...17..446M}. In this case the Peirce criterion is more appropriate than the widely use Chauvenet criterion \citep{1863QB145.C5.......}, because the dataset is very large and we need to reject more than one data point.

\section{Analysis}

The orbital analysis was performed on all $1536$ measurements reported to the MPC, plus $16$ positions obtained from our $5$ nights of Mauna Kea observations (see \tab{Astrometry}). We first identified the reference star catalog associated with each entry, applied the debiasing corrections as in \citet{2010Icar..210..158C}, and associated weights to each position with the approach described above.\\

\begin{deluxetable}{cccccccc}
\tabletypesize{\scriptsize}
\tablecaption{Astrometry\label{Astrometry}}
\tablewidth{0pt}
\tablehead{
\colhead{Date} & \colhead{$\alpha$ (J2000)} & \colhead{$\delta$ (J2000)} & \colhead{$R$} & \colhead{$\Delta\alpha_a$} & \colhead{$\Delta\delta_a$} & \colhead{$\Delta\alpha_c$} & \colhead{$\Delta\delta_c$}\\
\colhead{[UT]} & \colhead{[hh mm ss.sss]} & \colhead{[$\pm$dd~mm~ss.ss]} & & \colhead{[$\unsec$]} & \colhead{[$\unsec$]} & \colhead{[$\unsec$]} & \colhead{[$\unsec$]} 
}
\startdata
2011-08-01.413487 & 22 39 43.235 & +56 40 34.66 & 22.7 & 0.008 & 0.008 & 0.088 & 0.097 \\
2011-08-01.421289 & 22 39 41.662 & +56 40 39.46 & 22.5 & 0.008 & 0.007 & 0.044 & 0.066 \\
2011-08-01.430332 & 22 39 39.815 & +56 40 44.48 & 22.3 & 0.007 & 0.007 & 0.061 & 0.061 \\
2011-08-03.470563 & 22 38 21.505 & +56 33 07.30 & 22.3 & 0.006 & 0.006 & 0.035 & 0.026 \\
2011-08-03.474358 & 22 38 20.662 & +56 33 07.54 & 22.4 & 0.006 & 0.006 & 0.026 & 0.026 \\
2011-08-03.478086 & 22 38 19.833 & +56 33 07.55 & 22.3 & 0.006 & 0.006 & 0.035 & 0.035 \\
2011-08-03.486817 & 22 38 17.864 & +56 33 07.12 & 22.4 & 0.006 & 0.006 & 0.097 & 0.092 \\
2011-08-03.490571 & 22 38 17.020 & +56 33 06.76 & 23.0 & 0.006 & 0.006 & 0.026 & 0.040 \\
2011-08-03.494294 & 22 38 16.172 & +56 33 06.28 & 22.4 & 0.006 & 0.006 & 0.044 & 0.044 \\
2011-08-03.497970 & 22 38 15.337 & +56 33 05.77 & 23.0 & 0.007 & 0.006 & 0.079 & 0.061 \\
2011-08-07.534784 & 22 35 50.644 & +56 10 29.80 & 22.5 & 0.007 & 0.006 & 0.083 & 0.053 \\
2011-08-07.541449 & 22 35 49.300 & +56 10 25.06 & 22.6 & 0.007 & 0.006 & 0.097 & 0.110 \\
2011-08-07.548431 & 22 35 47.889 & +56 10 19.79 & 22.7 & 0.007 & 0.006 & 0.092 & 0.101 \\
2011-08-29.409194 & 22 23 48.426 & +51 35 39.03 & 23.2 & 0.008 & 0.007 & 0.035 & 0.035 \\
2011-08-29.417413 & 22 23 47.202 & +51 35 29.45 & 23.1 & 0.008 & 0.008 & 0.048 & 0.048 \\
2011-09-03.520191 & 22 21 17.166 & +49 38 16.08 & 23.0 & 0.007 & 0.008 & 0.140 & 0.110 \\
\enddata
\tablecomments{Astrometry, photometry and computed components of the error bar (from the astrometric solution and object centroid) for our observations of 2011~MD, referred to the PPMXL catalog. In addition to these error estimates, a catalog bias of $0.05\unsec$ was applied to each observation during our analysis.}
\end{deluxetable}

We used the orbital computation software \fo\footnote{\url{www.projectpluto.com/find\_orb.htm}} to compute a preliminary orbital solution, including all available observations, each weighted with its assigned error bar. Since the object is small, we took into account the possible effect of solar radiation pressure on the object, by allowing for an additional acceleration term in the radial direction. This dependence is parameterized with a single additional ``orbital'' element, the ratio between the average cross-sectional area of the object and its mass ($A/m$). It is important to point out here that the formal relation between the object's cross section and the corresponding radiation pressure acceleration actually involves a term dependent on the albedo (since each photon reflected by the asteroid surface transmits twice the momentum of an absorbed one); however, since the goal of the current section is only to parameterize the effect of radiation pressure on the orbit, we will here define the $A/m$ ratio as the one we would measure in the case of a perfect absorber, with zero albedo. The interplay of the albedo with the real cross section of the object will then be taken into account in the next section, when the $A/m$ will be used to estimate the physical properties of the object.

In addition to the effect of radiation pressure, it is also important to point out that 2011~MD came so close to our planet that higher-order multipole gravitational terms are significant. In this specific case the $J_2$ term turns out to be the dominant non-Newtonian term, a couple of orders of magnitude stronger than the radiation pressure effect. The software \fo~is capable of dealing with $J_2$, and it was included in our calculation. The next strongest term ($J_3$) is already much less important, because of the steeper radial dependency; the integrated acceleration caused by the $J_3$ multipole during the observed arc turns out to be $3$ orders of magnitude less than $J_2$, and about $30$ times less than the radiation pressure effect\footnote{The close approach of 2011~MD was so fast that the effective time of action of these multipole terms is tiny compared to the total observed arc, being of the order of $10^3 \un{s}$.}.\\

From this preliminary solution astrometric residuals were obtained. These residuals, together with each error bar, formed the basis for our rejection process, based on the Peirce criterion. For a description of the algorithm, and how it is applied to astrometric residuals (that are Rayleigh-distributed quantities) we refer the reader to the appendix of \citet{2012NewA...17..446M}. The algorithm rejected $144$ positions, out of a total set of $1543$. Most of the rejections ($96$ positions) came from the single station I27 (Barred Owl Observatory), and they would have dominated the solutions if not properly rejected and deweighted.\\

After the rejections, we re-computed a full orbital solution, including again the effects of $J_2$ and radiation pressure. This solution corresponds to a statistically significant detection of solar radiation pressure acting on the object. The best-fit value for $A/m $ is $0.32 \times 10^{-3} \un{m}^2\un{kg}^{-1}$, comparable with the value presented in \citet{2012NewA...17..446M} for 2009~BD, an object of similar size.

We then computed an error bar for this value by creating Monte Carlo samples of the outlier-rejected observation set, with the addition of Gaussian noise proportional to the error bar of each observation. Our resulting $1\sigma$ estimate is $A/m = (0.32 \pm 0.06) \times 10^{-3} \un{m}^2\un{kg}^{-1}$, corresponding to a $5 \sigma$ detection of a non-zero value.

\section{Discussion}

The $A/m$ value reported above can be used to extract useful physical information on the object. 
It is important to point out that 2011~MD is in an extremely Earth-like orbit, with low $\Delta v$ with respect to our planet. It is also a small object, with $H \sim 28$, corresponding to a diameter of a few meters (depending on the albedo). It is therefore tempting to assume that it might not be a natural object, but rather a piece of debris of man-made origin (such as an upper stage of a rocket)\\

To prove beyond doubt the natural nature of 2011~MD it is possible to use the $A/m$ value to put constraints on the density of the object. This can be done only under a series of assumptions, that will result in a correspondingly larger error bar in the density estimate.

The first and most relevant assumption is about the albedo. Unfortunately no direct or indirect estimate of the albedo is available for 2011~MD, since no thermal data were obtained during the fly-by. No color information is also available, making it impossible to restrict the range of likely values based on its spectral class. We are therefore forced to assume the broadest possible distribution for this parameter, and convolve it with the other measured quantities to compute an appropriate error bar for our density estimate. We choose to assume a probability distribution for the albedo based on the data presented by \citet{2014ApJ...784..110M}, which are specifically restricted to small NEAs and therefore appropriate for our analysis.

The second value we need for a density estimate is an accurate absolute magnitude for the object, that combined with the albedo will give us an estimate of the size. Again, no well-calibrated photometry is available for 2011~MD, but we can provide an estimate based on the photometry values reported to the MPC together with the astrometric positions.
The nominal absolute magnitude is around $H=28.0$, corresponding to an approximate size of about $10\un{m}$ assuming a typical NEO albedo of $12\%$. 
To attach an error bar to this value we need to take into account both the statistical error of the determination and the rotational variability of the object. The second is the dominant factor in this case; useful information on the amplitude can be obtained from various sources, including the set of $968$ observations reported by station I27 and already discussed for their astrometric relevance, or various analyses published online\footnote{An example of a folded high-SNR lightcurve obtained around close approach is available at \url{www.nmt.edu/\~{}bryan/research/work/mro\_images/k11m00d/}.}. From them it is possible to extract a time interval between consecutive minima of $\sim 697 \un{s}$, with a peak to peak amplitude of about $0.85$ magnitudes. However, these estimates were obtained at a high phase angle ($\alpha \sim 60\undeg$), and they are not representative of the true zero-phase lightcurve of 2011~MD. We can empirically correct the observed amplitude to a zero-phase amplitude using the approach of \citet{1990A&A...231..548Z}, obtaining a full amplitude of $\sim 0.45$ magnitudes\footnote{The phase factors listed by \citet{1990A&A...231..548Z} are dependent on the asteroid spectral type. To follow the most conservative approach we used the conversion factor for M-type asteroids, which corresponds to the largest zero-phase amplitude.}. To maintain our conservative approach, we are therefore assuming an error bar of $\pm 0.3$ on our $H$ value, which includes both the uncertainty in magnitude and in the actual shape of 2011~MD.\\

\begin{figure}[t]
\epsscale{1.0}
\plotone{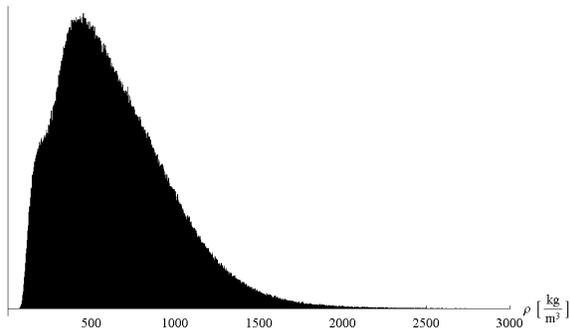}
\caption{Probability distribution (with arbitrary normalization) of the density for 2011~MD.\label{Histogram}}
\end{figure}

\begin{figure}[t]
\epsscale{1.0}
\plotone{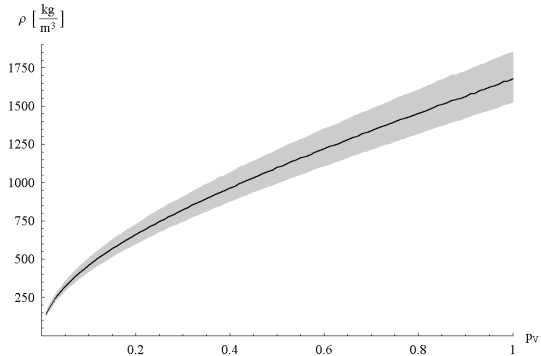}
\caption{Density of 2011~MD as a function of the assumed albedo, and corresponding $1\sigma$ band resulting from the uncertainty of the other parameters.\label{Plot}}
\end{figure}

Under these assumptions, it is possible to estimate the probability distribution for the density of our object, convolving the distributions of $A/m$, the albedo and the absolute magnitude, and following the relations derived by \citet{2000A&A...362..746V} for the assumption of a spherical Lambertian reflector. The result is shown in \fig{Histogram}; it is immediately evident that the average value is low, well below water density. In particular, we obtain an average bulk density of $\rho = 640 \unrho$, with a $1\sigma$ range (intended here as $68\%$ of the distribution) that extends from $310 \unrho$ to $970 \unrho$, and a $3\sigma$ interval of $100 \unrho$ to $2200 \unrho$. The dependency of the density from the assumed albedo is better clarified in \fig{Plot}. 


This value, while surprisingly low, is in good agreement with the estimates presented by \citet{2012NewA...17..446M, 2013Icar..226..251M} for 2009~BD and 2012~LA, two other objects of similar size, and analyzed under similar assumptions (although using a more general albedo distribution). This peculiar behavior seems to point to either a general low bulk density (and likely very high porosity) of these small objects, or an anomalously high albedo, which needs to be assumed at the level of $p_V\sim0.5$ to be compatible with typical densities of even the lightest major asteroidal bodies like (253) Mathilde \citep{1999Icar..140....3V}, or with larger NEOs like (101955) Bennu \citep{2014Icar..235....5C}, both of which have measured densities of $\rho \sim 1300 \unrho$ but are instead known to be extremely dark ($p_V\sim0.04$).

Both these interpretations, if confirmed, can have significant implications on the estimate of hazard from impact of very small bodies, which actually represent the most likely population of impactors, at least on the short term. Furthermore, since 2011~MD is currently considered a prime target for a possible ARRM, a proper characterization of its physical nature (especially size and mass) is essential for the definition of an appropriate mission profile to the object.

It is also important to point out that the nominal density value we obtained, while low, is still well above the expected bulk density for man-made objects. A typical upper stage of a rocket, while being mostly made of metal, is generally a hollow cylindrical shell, and its bulk density is typically between $20 \unrho$ to $50 \unrho$, about an order of magnitude less than our estimate. We can therefore at least exclude the artificial nature of the target, an important information in case an ARRM mission plans to reach and retrieve it for further study.

\section{Conclusions}

From the observational data presented above we obtained a statistically significant detection of the action of radiation pressure on the small object 2011~MD, based on a relatively short observational arc (only $73$ days). To our knowledge this is the first detection of a non-gravitational effect on a natural object observed during a single close encounter with our planet, and shows the value of high-precision astrometry and of a proper statistical treatment of astrometric data. 
It is worth noting that the data used in this work are only optical, without any radar detection.

The most relevant scientific result of this work is the low density value $\rho = (640\pm330) \unrho$ obtained for 2011~MD under assumptions of a typical albedo probability distribution. While well above typical bunk densities of man-made objects, it still unexpectedly low for a natural object, and would imply either an extremely high bulk porosity, or an estimate biased by an unusually high albedo, and therefore a significantly smaller diameter (about $5\un{m}$ if we assume $p_V\sim0.5$). Both these interpretations can have significant implications in terms of impact hazard from small objects, but also in light of a possible ARRM to this target or to others with comparable properties.



\acknowledgments

Our observations of 2011~MD were funded by grant AST 0709500 from the U.S. National Science Foundation.

The authors would like to thank Bill Gray for developing the software \fo, which made most of this analysis possible, and for the fruitful e-mail interaction leading to updates which made it more accurate and effective.

The authors wish to recognize and acknowledge the very significant cultural role and reverence that the summit of Mauna Kea has always had within the indigenous Hawaiian community. We are most fortunate to have the opportunity to conduct observations from this sacred mountain.



{\it Facilities:} \facility{UH:2.2m}.

\end{document}